# A Conceptual DFT Approach Towards Developing New QSTR Models†

**P. K. Chattaraj[a,*], D. R. Roy[a], S. Giri[a], S. Mukherjee[a], V. Subramanian[b,*], P. Bultinck[c,*] and S. Van Damme[c]**

[a]*Department of Chemistry, Indian Institute of Technology, Kharagpur 721 302, India*

[b]*Chemical Laboratory, Central Leather Research Institute, Adyar, Chennai 600 020, India*

[c]*Department of Inorganic and Physical Chemistry, Ghent University, Krijgslaan 281, B-9000 Gent, Belgium*

**Abstract**

Quantitative-structure-toxicity-relationship (QSTR) models are developed for predicting the toxicity ($pIGC_{50}$) of 252 aliphatic compounds on *Tetrahymena pyriformis*. The single parameter models with a simple molecular descriptor, the number of atoms in the molecule, provide unbelievable results. Better QSTR models with two parameters result when global electrophilicity is used as the second descriptor. In order to tackle both charge- and frontier-controlled reactions the importance of the local electro (nucleo) philicities and atomic charges is also analyzed. Best possible three parameter QSTR models are prescribed.

†This paper is dedicated to Professor Robert G. Parr, the father of the conceptual DFT, on the occasion of his eighty-fifth birthday.

***Author for correspondence**: pkc@chem.iitkgp.ernet.in

## 1. Introduction

Ever since the power of Quantitative- structure- activity- relationship (QSAR) based techniques has been highlighted, several descriptors have been proposed from time to time in developing QSAR models[1-8] for understanding various aspects of pharmacological sciences including drug design and the possible ecotoxicological characteristics of the drug molecules. Specific quantitative structure toxicity relationship (QSTR) models have also been developed. In these studies the toxicity of various chemicals have been understood via corresponding molecular structures. An extensive research has been carried out[9-16] in understanding the toxicological effects of several aliphatic compounds on ciliated protozoa called *Tetrahymena pyriformis.* Both European Union and U.S. Environmental Protection Agency require reliable toxicity data set for various classes of living systems like primary producers, invertebrates and vertebrates. This information is used for QSAR/QSTR as well as regulatory purposes. The ciliated protozoa, *Tetrahymena pyriformis* has been considered to be ideal for the associated laboratory research. In this ciliate species, diverse endpoints can be used to originate the cytotoxic effects and xenobiotics. Experimental determination of toxicological and biochemical endpoints is a difficult task. Hence, QSAR/QSTR modeling of the toxicity of aliphatic compounds on the *T. pyriformis* is of vital importance in investigation of its toxicity in terms of its inhibitory growth concentration (IGC). A multitude of QSTR models exist which analyze the associated toxicity behavior. Quantum chemical descriptors[17-20] have also been used for this purpose and they have been proved to be versatile and reliable.

Toxicity analyses of a diverse class of systems have been carried out using conceptual density functional theory (DFT) based reactivity/selectivity descriptors. Possibility of electron transfer between a toxic molecule and a biosystem has been considered to be one of the major reasons of toxic behavior of these molecules. Accordingly the related descriptors like electron affinity, ionization potential, planarity, electrophilicity etc. have been turned out to be useful QSTR descriptors. Experimental toxicity values of a wide variety of polyaromatic hydrocarbons like polychlorinated dibenzofurans (PCDFs), polychlorinated biphenyls (PCBs), polychlorinated dibenzo-p-dioxins (PCDDs) and chlorophenols (CP), several aliphatic and aromatic toxic molecules have been shown to correlate very well[20-30] with the corresponding toxicity values estimated using various conceptual DFT descriptors especially global and local electrophilicities.[31-33]

Several researchers[9-16] have studied the toxicological behavior of various compounds on *Tetrahymena pyriformis*. They have highlighted the importance of the studies as well as the possibility of constructing a large number of QSTR models with a varied range of success and the difficulty in computation. A state-of-the art QSTR model has been developed by Schultz et al.[13] Toxicity of a large number of aliphatic compounds on *Tetrahymena pyriformis* has been studied[13] through QSTR models developed[13] in terms of logP and the lowest unoccupied molecular orbital energy ($E_{LUMO}$) whereas the effect of several aromatic compounds on the same system has been analyzed[14] in terms of log P, $E_{LUMO}$ and maximum acceptor superdelocalizability ($A_{max}$). In both cases the models are found to be robust. We have shown[20,25] that global and local electrophilicities are useful descriptors of toxicity prediction. In the present work we propose to develop

QSTR models for toxicity of several aliphatic compounds on *Tetrahymena pyriformis,* using a simple descriptor, viz. the number of atoms present in the molecule, which can be obtained without even touching a computer. Section 2 provides the theoretical background whereas the computational details are provided in section 3. Results and discussion are provided in section 4 and finally, section 5 contains some concluding remarks.

## 2. Theoretical Background

We consider the number of atoms in a molecule to be a valid descriptor of its toxic nature. For a given group of molecules the number of electrons (N) is expected to scale as the number of atoms present ($N_a$). Molecules with larger $N_a$ values are supposed to have larger molecular weights implying larger log P values. That in turn will provide larger toxicity values. For simplicity we consider the number of carbon atoms ($N_C$) as the variable and for the set of molecules with a constant $N_C$ we may choose the number of non-hydrogenic atoms ($N_{NH}$) as the descriptor. Related descriptors have been used in the past.[34]

In order to have a complete analysis we also check the nature of the model where electrophilicity ($\omega$) is used as an additional descriptor, which has been shown[20,25] to be a reliable descriptor of biological activity[19] and toxicity.[20,25] The electrophilicity is defined as[31, 32]

$$\omega = \frac{\mu^2}{2\eta} \qquad (1)$$

where $\mu = -\frac{I+A}{2}$ and $\eta = \frac{I-A}{2}$ are the electronic chemical potential and hardness respectively. *I* and *A* being the ionization potential and electron affinity respectively.

It has also been shown[20,25] that apart from global electrophilic power the local electro (nucleo) philicity is important in understanding the possible charge transfer between a toxin and a receptor. The philicity at an atom *k* of the molecule is defined as[33]

$$\omega_k^{\alpha} = \omega . f_k^{\alpha} \qquad (2)$$

where $\left\{ f_k^{\alpha} \right\}$ are the condensed-to-atom-*k* Fukui functions calculated in terms of the electronic population $q_k$ and $\alpha$ = +, - and 0 refers to nucleophilic, electrophilic and radical attacks respectively. The condensed Fukui functions are given by[35]

$$f_k^{+} = q_k(N+1) - q_k(N) \qquad (3a)$$

$$f_k^{-} = q_k(N) - q_k(N-1) \qquad (3b)$$

$$f_k^{0} = [q_k(N+1) - q_k(N-1)]/2 \qquad (3c)$$

Since the Fukui function based descriptors are ideally suited for soft-soft-frontier-controlled reactions and the atomic charges ($Q_k$) in a molecule are known to be appropriate local descriptors in analyzing essentially charged-controlled reactions between a hard nucleophile and a hard electrophile[36-38] we also consider the latter in our analysis.

Comparing the electronegativity values of 13 sets of aliphatic compounds whose toxic nature towards *Tetrahymena pyriformis* is known,[9-16] with those of various nucleic acid bases (adenine, thymine, guanine, cytosine and urasil) and DNA base pairs (GCWC and ATH) it was observed[20] that there are nine groups of electron acceptors (saturated alcohols, diols, halogenated alcohols, mono- and di- esters, carboxylic and halogenated acids, aldehydes and ketones) and four groups of electron donors (unsaturated alcohols, $\alpha$ -acetylinic alcohols, amino alcohols and amines). For the former group $\omega_{\max}^{+}$ and for the latter group $\omega_{\max}^{-}$ are considered to be[20] appropriate descriptors where $\omega_{\max}^{\alpha}$ refers to the $\omega_{k}^{\alpha}$ value at the site where it is maximum. For the hard interactions, $Q_{k}^{\max}$ is considered to be the proper descriptor where *k* is the site with the maximum value of the magnitude of the charge (positive for the acceptors and negative for the donors).

## 3. Computational Details

Geometries of all the 252 aliphatic molecules (acceptors-171, donors-81) corresponding to the 13 groups are optimized at the Hatree-Fock level with 6-311G* basis set using the *Gaussian 03*[39] program. These molecules were tested before[20, 25] for correlating their experimental log($IGC_{50}^{-1}$) values[10] against *Tetrahymena pyriformis* with the corresponding values calculated in terms of global and local electrophilicities.

Equations (1)-(3) are used to calculate the global and local electrophilicities. Necessary population and charges are calculated using the natural population analysis (NPA) scheme. Single point calculations are done for the (N±1) - electron systems with the N-electron molecule geometry.

In the statistical analysis, the systematic search is performed to determine statistically significant relationships between the toxicity and a selection of one, two or three descriptors out of the six available descriptors ($N_C$, $N_{NH}$, $\omega$, $\omega_{\max}^{+}$, $\omega_{\max}^{-}$ and $Q_{k}^{\max}$). The analysis is performed using in-house software. In order to minimize the effect of multicollinearity and to avoid redundancy, the descriptor set is first pre-evaluated with unsupervised forward selection. This selection is a variable elimination technique where variables are physically removed from the data set. Variables are eliminated for two reasons. First, they are eliminated if they have a small variance, below some threshold value. The second reason for variable removal is the existence of redundancy (exact linear dependencies between subsets of the variables) and multicollinearity (high multiple correlations between subsets of the variables) in QSAR data sets. Multicollinearity and redundancy may result in highly unstable estimates for regression coefficients, because their values may change enormously when variables are added or deleted to the regression. Both these features are assessed by inspecting the multiple correlations within the relevant subsets of descriptors. For a detailed overview of the UFS algorithm we refer to references 40 and 41. As regression technique the multiple linear regression is preferred over principal component regression or partial least squares regression, because of the ease of interpretation of the outcome. The output provided in the paper gives the most promising 3-parameter model for each dataset. The following statistical criteria of the models are noted: R, R-square, adjusted R-square and the standard error of the estimate are measures to confirm a good fit of the data to the regression line. Internal validation is conducted with leave-one-out cross-validation and is given by $Q^2$. The significance of this value is estimated by Y-randomisation.

A mechanistic interpretation can be deduced from the output, by using the coefficients (b). These are descriptors calculated from scaled data values. This gives the opportunity to evaluate the descriptors in relation to each other. Outliers are detected graphically in the regression diagnostic plot.

Each statistical analysis is preceded by an analysis of the dataset. A graphical analysis of the residuals (residual plot, normal probability plot and regression plot) permits the user to confirm if the dataset is suitable for a multiple linear regression.

## 4. Results and Discussion

Experimental and calculated $pIGC_{50}$ values along with various descriptors are presented in Tables 1 and 2 for the electron acceptor and electron donor molecules respectively. Table 3 presents the corresponding regression equations associated with one ($N_C$)- and two ($N_C, \omega$)- parameter models. The single parameter model with a simple descriptor like the number of carbon atoms provides good estimates of toxicity in most cases. The exceptions are carboxylic acids, halogenated acids, amino alcohols and amines. Situation improves drastically in all cases except for amines where the information of charge is also important. The constant terms in the two-parameter models are not always significant. Figure 1 presents the plots of experimental $pIGC_{50}$ versus calculated $pIGC_{50}$ values for the a) acceptor set ($R^2$=0.9283, $R^2_{CV}$=0.9279, $R^2_{Adj}$=0.9265) and b) donor set ($R^2$=0.8284, $R^2_{CV}$=0.8262, $R^2_{Adj}$=0.8156) which authenticates the efficacy of these regression models for QSTR. It is also important to note that the slopes of these plots are unity and the intercepts are very close to zero, as expected. Although the predicted toxicity trend is satisfactory when compared with the observed one, for the individual outlier molecules it is difficult to provide with a rationale a priori. It may be noted that the calculated $pIGC_{50}$ values plotted in Figure 1 are obtained through different regression models for 13 different sets of molecules. In each set the molecules of similar chemical behavior are included. In case we take all the molecules together the following regression equations are obtained:

$$pIGC_{50} = 0.2789 \times N_C - 2.2484$$

$$R = 0.805;\ SD = 0.551;\ N = 252 \qquad (4a)$$

$$pIGC_{50} = 0.2838 \times N_C + 0.6415 \times \omega \ -2.7888$$

$$R = 0.812;\ SD = 0.542;\ N = 252 \qquad (4b)$$

It may be noted that for a diverse class of chemical compounds $N_C$ and ω may still be considered to be useful descriptors. Corresponding plots are provided in Figure 2 (a, b). The correlation improves further in case a couple of sets are removed as was done by Schultz et al.[13] For the shake of completeness we also include the plot of the experimental toxicity with log P for the same 252 molecules (Figure 2c). The correlation is comparable to that obtained in Figure 2a which is expected because of the inter correlation between log P and $N_C$ (Figure 2d). It may be noted that unlike log P, $N_C$ does not require any software (instrument) to compute it (determine it experimentally).

Now we try to develop the best possible three-parameter models (with the highest $R^2$ value) by selecting all possible combinations of 1, 2 and 3 parameters chosen out of six possibilities, viz., $N_C$, $N_{NH}$, $\omega$, $\omega^{+}_{\max}$, $\omega^{-}_{\max}$ and $Q_k^{\max}$. Various plots and the model summary (not shown here) associated with each set reveal that it is possible to develop beautiful QSTR models using three parameters. They are presented sequentially. Now we will investigate for each set of molecules whether we can obtain three-parameter models which behave better than the already obtained one-and two parameter models. The best three-parameter model is chosen by investigating all possible combinations of 3 parameters out of a list of six available descriptors. The best three-parameter model is chosen based on the highest $R^2$ value. To compare three-parameter models with one-and two parameter models, we cannot use the $R^2$ value. The value of $R^2$ can generally be increased by adding additional descriptor variables to the model, even if the added variable does not contribute to reduce the unexplained variance of the dependent variable. This can be avoided by using another statistical parameter – the so-called adjusted $R^2$ ($R^2_{adj}$).

Before any regression analysis can take place, we have to check the data set on a few principal assumptions. These assumptions justify the use of linear regression models for purposes of prediction: Independence, normality and linearity. If any of these assumptions is violated, then the insights yielded by a regression model may be inefficient or seriously biased or misleading. The characteristics of the data sets are checked visually. For a detailed overview of these assumptions and their visualization, the reader is referred to the reference 42. For the purpose of this article it is sufficient to look at the graphics (a), (c) and (d) for each of the data sets. The graphic (a) has to be a scattered plot of points around zero and the graphics (c) and (d) have to be a straight line of points through the origin. As one can see, for each of the thirteen data sets, these characteristics are fulfilled.

The performance of the multiple linear regression is summarized in a few statistical parameters. The most important ones for this purpose are $R^2_{adj}$, the standard error of the estimate, $Q^2$ and the F-ratio. Each of these terms is explained in the previous section and their behavior can be found in statistical literature. If we encounter a model which does not behave well for one of these parameters, the model has to be rejected. Based on these four statistical parameters, each of the models can be accepted as statistically significant models.

The fact that each of these 3- parameter models is statistically significant does not mean that they behave better then the corresponding 1-or 2-parameter models. In view of the behavior of these models one has to look at three parameters. First of all, as mentioned before, one has to compare the adjusted $R^2$ of the three-parameter model with those of the one-or two-parameter models. All of the three-parameter models concerning $R^2$ behave better than the corresponding one-or two-parameter models, except the model for the set of alpha-acetylenic alcohols, which behave worse than the two-parameter model.

| Molecule set | Two parameter model | Three parameter model |
|---|---|---|
| Alpha-acetylenic alcohols | 0.8085 | 0.7872 |

For the following three-parameter models the $R^2$ increases only slightly by adding one descriptor.

| Molecule set | Two parameter model | Three parameter model |
| --- | --- | --- |
| Saturated Alcohols | 0.9261 | 0.9267 |
| Diester | 0.9245 | 0.9272 |

Since it is better to have a model with as least as possible descriptors, according the Principle of Parsimony,[44] we prefer for these sets the two-parameter models.

Three parameter models are only better then the two-parameter models, if the three parameters used are statistically significant. The in-house built statistical software conducts this test of significance with a Students t-test. The following models are assigned containing non-significant descriptors:

| Molecule set | Non significant parameters |
| --- | --- |
| Halogenated Acids | $N_C$<br>$Q_k^{max}$ |
| Aldehydes | $\omega$ |
| Amino Alcohols | $N_C$ |

The remaining models can be used for predictive purposes, only if they do not fail on the Y-randomisation test. As can be seen on the graphs (e), this concerns only the three-parameter models of the aliphatic amines, the unsaturated alcohols, the monoesters and the carboxylic acids. The three parameter models for these sets are

| Molecule set | | | |
| --- | --- | --- | --- |
| Carboxylic acid | $\omega$ | $N_C$ | $Q_k^{max}$ |
| Monoester | $N_{NH}$ | $\omega$ | $\omega_{max}^{+}$ |
| Unsaturated Alcohol | $\omega$ | $\omega_{max}^{-}$ | $N_C$ |
| Aliphatic amines | $\omega_{\max}^{-}$ | $\omega$ | $Q_k^{\max}$ |

The fact that the remaining models of the halogenated alcohols, the diols and the ketones do not provide good Y-randomisation test results might be originating from the small number of molecules (respectively 11, 10 and 15 ) for the number of descriptors used.[45]

It is important to note that $\omega_{\max}^{+}$ and $\omega_{\max}^{-}$ respectively appear in the electron acceptor and donor sets, as expected. Except for the set of aliphatic amines the atom number is a valuable descriptor. The global electrophilicity is also a reliable descriptor in most cases.

In order to check the efficacy of the present QSTR models vis-à-vis other existing popular models we make a comparative analysis of our work with that reported by Schultz et al[13] which may be considered to be by far the best available till date. To compare the quality of regression we analyze our Figure 2 and Eq.(4) with their Figures 1 & 2 and Eq.(1). As opposed to the regression presented for only neutral narcotics,[13] our mixture of a wide class of compounds shows a reasonable correlation. Like our correlations for 13 separate groups (both polar and nonpolar) as presented in Table 3, their one parameter models for nonpolar molecules (Table 1 of Ref.13) and two parameter models for 7 selected classes (Table 2 of Ref.13) we notice the following: a) For one parameter models their slopes are positive and intercepts are negative for all 5 sets and are of same orders of magnitude. For one parameter models our slopes are positive and intercepts are negative for all 13 sets and are of same orders of magnitude. Moreover, obtaining log P is much more difficult than counting $N_C$., b) For the two parameter models their coefficients for log P are always positive as in our case the coefficients for $N_C$ are also always positive. In their models the coefficients for $E_{LUMO}$ and the intercept are positive in some cases and negative in other cases as is the case with our coefficients of ω and the intercept., c) There is a good correlation[43] between log P and $N_C$. It is expected that our works would be complimentary to their work., d) The problems they faced in modeling carboxylic acids, amino alcohols, α-halo-activated compounds etc. may be better tackled using these descriptors or a combination of them (log P ($N_C$), $E_{LUMO}$, ω)., e) Importance of charge as the descriptor in modeling amines is now understood, f) Macroscopic descriptors like log P or $N_C$ would be useful for a broad spectrum of systems. However, electronic descriptors like ω would be useful when systems with similar electronic environment are analyzed. They would be specially useful when molecules will have nearly identical or identical log P ($N_C$) values. For molecules with similar electronic environment local (or group) electrophilicity would highlight the importance of the site (group) especially for the toxic behavior.

## 5. Concluding Remarks

The number of atoms in a molecule can provide important insights into its possible toxic behavior. It can be used as a molecular descriptor for predicting $pIGC_{50}$ values of various aliphatic compounds against the ciliate *Tetrahymena pyriformis*. The situation improves further when electrophilicity is used as an additional descriptor. Local electro (nucleo) philicity and atomic charges are also considered to take care of local soft-soft and hard-hard interactions, which resulted in robust three parameter QSTR models.

## Acknowledgements

We thank BRNS, Mumbai for financial assistance and unknown referees for constructive criticism. One of us (P.B.) acknowledges the Fund for Scientific Research-Flanders (FWO-Vlaanderen) for continuous support for his group.

Table Captions:

**Table 1**: Electrophilicity (ω), maximum atomic charge ( $Q_k^{\max}$ ), number of carbon atoms (Nc), log P along with the experimental and calculated values of log ($IGC_{50}^{-1}$) for the complete set of aliphatic acceptor compounds with *Tetrahymena pyriformis*.

**Table 2**: Electrophilicity (ω), maximum atomic charge ( $Q_k^{\max}$ ), number of carbon atoms (Nc), log P along with the experimental and calculated values of log ($IGC_{50}^{-1}$) for the complete set of aliphatic donor compounds with *Tetrahymena pyriformis*.

**Table 3.** Regression models for different groups of aliphatic compounds for estimating their toxicity towards *Tetrahymena pyriformis*

Figure Captions:

**Figure 1.** Observed versus calculated $pIGC_{50}$ values using two-parameter ($\omega$, $N_C$) regression models for the (a) Complete set of aliphatic electron acceptors and (b) Complete set of aliphatic electron donors. Please see the text for details.

**Figure 2.** Observed $pIGC_{50}$ versus the a) Number of carbon atoms ($N_C$), b) Calculated $pIGC_{50}$ values using two-parameter ($\omega$, $N_C$) regression model and (c) log P along with the (d) inter-correlation between log P and $N_C$ for the complete set of 252 aliphatic compounds.

**Table 1**: Electrophilicity (ω), maximum atomic charge ( $Q_k^{max}$ ), number of carbon atoms (Nc), log P along with the experimental and calculated values of log ($IGC_{50}^{-1}$) for the complete set of aliphatic acceptor compounds with *Tetrahymena pyriformis*.

| Molecules | ω | $Q_k^{max}$ | Nc | log P* | pIGC50 | | |
|---|---|---|---|---|---|---|---|
| | | | | | Expt.* | Calc. (Nc) | Calc. (Nc, ω) |
| ***Diols*** | | | | | | | |
| (+/-)-1,2-butanediol | 0.8999 | 0.4652 | 4 | -0.53 | -2.0482 | -2.2868 | -2.1479 |
| (+/-)-1,3-butanediol | 0.9286 | 0.4488 | 4 | -1.38 | -2.3013 | -2.2868 | -2.5044 |
| 1,4-butanediol | 0.8915 | 0.4492 | 4 | -0.83 | -2.2365 | -2.2868 | -2.0435 |
| 1,2-pentanediol | 0.8907 | 0.4652 | 5 | 0.00 | -1.6269 | -1.8371 | -1.6782 |
| 1,5-pentanediol | 0.9132 | 0.4487 | 5 | -0.64 | -1.9344 | -1.8371 | -1.9577 |
| 2-methyl-2,4-pentanediol | 0.9200 | 0.4463 | 6 | -0.68 | -1.9531 | -1.3874 | -1.6868 |
| (+/-)-1,2-hexanediol | 0.8887 | 0.4652 | 6 | 0.53 | -1.2669 | -1.3874 | -1.2979 |
| 1,6-hexanediol | 0.9027 | 0.4487 | 6 | -0.11 | -1.4946 | -1.3874 | -1.4719 |
| 1,2-decanediol | 0.8640 | 0.4651 | 10 | 2.64 | 0.7640 | 0.4113 | 0.4305 |
| 1,10-decanediol | 0.8597 | 0.4484 | 10 | 2.01 | 0.2240 | 0.4113 | 0.4839 |
| ***Halogenated Alcohol*** | | | | | | | |
| 2-Bromoethanol | 0.9418 | 0.4575 | 2 | 0.18 | -0.8457 | -1.3706 | -0.9446 |
| 2-Chloroethanol | 1.0417 | 0.4578 | 2 | -0.06 | -1.4174 | -1.3706 | -1.5727 |
| 1-Chloro-2-propanol | 1.0170 | 0.4549 | 3 | 0.14 | -1.492 | -1.0434 | -1.2191 |
| 3-Chloro-1-propanol | 1.0101 | 0.4525 | 3 | 0.50 | -1.3992 | -1.0434 | -1.1758 |
| 4-Chloro-1-butanol | 0.9570 | 0.4514 | 4 | 0.85 | -0.7594 | -0.7163 | -0.6437 |
| 3-Chloro-2,2-dimethyl-1-propanol | 0.9843 | 0.4553 | 5 | 0.81 | -0.7822 | -0.3892 | -0.6171 |
| 6-Chloro-1-hexanol | 0.9417 | 0.4497 | 6 | 1.59 | -0.2726 | -0.0621 | -0.151 |
| 8-Chloro-1-octanol | 0.9278 | 0.4490 | 8 | 2.65 | 0.4878 | 0.5921 | 0.3329 |
| 6-Bromo-1-hexanol | 0.8636 | 0.4497 | 6 | 1.73 | 0.0074 | -0.0621 | 0.3399 |
| 8-Bromo-1-octanol | 0.8559 | 0.4490 | 8 | 2.79 | 1.0424 | 0.5921 | 0.7848 |
| 2,3-Dibromopropanol | 0.9902 | 0.4599 | 3 | 0.63 | -0.4861 | -1.0434 | -1.0507 |
| ***Saturated Alcohol*** | | | | | | | |
| methyl alcohol | 0.9485 | 0.4440 | 1 | -0.77 | -2.6656 | -2.6657 | -2.6755 |
| ethyl alcohol | 0.9186 | 0.4481 | 2 | -0.31 | -1.9912 | -2.2513 | -2.2761 |
| 1-propanol | 0.8979 | 0.4485 | 3 | 0.25 | -1.7464 | -1.8369 | -1.8685 |
| 2-propanol | 0.9500 | 0.4548 | 3 | 0.05 | -1.8819 | -1.8369 | -1.822 |
| 1-butanol | 0.8960 | 0.4484 | 4 | 0.88 | -1.4306 | -1.4225 | -1.4441 |
| (+/-)-2-butanol | 0.9227 | 0.4480 | 4 | 0.61 | -1.542 | -1.4225 | -1.4202 |
| 2-methyl-1-propanol | 0.9066 | 0.4501 | 4 | 0.76 | -1.3724 | -1.4225 | -1.4346 |
| 2-pentanol | 0.9045 | 0.4479 | 5 | 1.19 | -1.1596 | -1.0081 | -1.0104 |
| 3-pentanol | 0.8945 | 0.4569 | 5 | 1.21 | -1.2437 | -1.0081 | -1.0193 |
| 3-methyl-2-butanol | 0.8935 | 0.4482 | 5 | 1.28 | -0.9959 | -1.0081 | -1.0202 |
| tert-amylalcohol | 0.9354 | 0.4459 | 5 | 0.89 | -1.1729 | -1.0081 | -0.9828 |
| 2-methyl-1-butanol | 0.9034 | 0.4502 | 5 | 1.22 | -0.9528 | -1.0081 | -1.0114 |
| 3-methyl-1-butanol | 0.9218 | 0.4481 | 5 | 1.16 | -1.0359 | -1.0081 | -0.9949 |
| 2,2-dimethyl-1-propanol | 0.9416 | 0.4516 | 4 | 1.31 | -0.8702 | -1.4225 | -1.4034 |
| 2-methyl-2-propanol | 0.9560 | 0.4446 | 4 | 0.35 | -1.7911 | -1.4225 | -1.3905 |
| 1-hexanol | 0.8955 | 0.4484 | 6 | 2.03 | -0.3789 | -0.5936 | -0.5923 |
| 3,3-dimethyl-1-butanol | 0.9357 | 0.4483 | 5 | 1.62 | -0.7368 | -1.0081 | -0.9825 |
| 4-methyl-1-pentanol | 0.9354 | 0.4484 | 6 | 1.75 | -0.6372 | -0.5936 | -0.5567 |
| 1-heptanol | 0.8958 | 0.4484 | 7 | 2.72 | 0.1050 | -0.1792 | -0.1659 |
| 2,4-dimethyl-3-pentanol | 0.8519 | 0.4525 | 7 | 1.93 | -0.7052 | -0.1792 | -0.2051 |
| 1-octanol | 0.8769 | 0.4483 | 8 | 3.00 | 0.5827 | 0.2352 | 0.2433 |
| 2-octanol | 0.8779 | 0.4479 | 8 | 2.90 | 0.0011 | 0.2352 | 0.2442 |
| 3-octanol | 0.8560 | 0.4511 | 8 | 2.72 | 0.0309 | 0.2352 | 0.2247 |
| 1-nonanol | 0.8560 | 0.4483 | 9 | 3.77 | 0.8551 | 0.6496 | 0.6508 |
| 2-nonanol | 0.8658 | 0.4479 | 9 | 3.25 | 0.6183 | 0.6496 | 0.6595 |
| 3-ethyl-2,2-dimethyl-3-pentanol | 0.8221 | 0.4483 | 9 | 2.86 | -0.1691 | 0.6496 | 0.6205 |
| 1-decanol | 0.8387 | 0.4483 | 10 | 4.57 | 1.3354 | 1.064 | 1.0614 |
| (+/-)-4-decanol | 0.8182 | 0.4512 | 10 | 3.78 | 0.8499 | 1.064 | 1.0431 |
| 3,7-dimethyl-3-octanol | 0.8658 | 0.4450 | 10 | 3.52 | 0.3404 | 1.064 | 1.0856 |
| 1-undecanol | 0.8248 | 0.4483 | 10 | 4.53 | 1.9547 | 1.064 | 1.049 |
| 1-dodecanol | 0.8132 | 0.4483 | 12 | 5.13 | 2.1612 | 1.8928 | 1.8909 |
| 1-tridecanol | 0.8035 | 0.4483 | 13 | 5.58 | 2.4497 | 2.3072 | 2.3083 |
| ***Carboxylic Acid*** | | | | | | | |
| Propionic acid | 0.9901 | 0.9780 | 3 | 0.33 | -0.5123 | -0.6331 | -0.431 |
| Butyric acid | 1.0051 | 0.9840 | 4 | 0.79 | -0.5720 | -0.5216 | -0.4788 |
| Valeric acid | 0.9840 | 0.9851 | 5 | 1.39 | -0.2674 | -0.41 | -0.3301 |
| Hexanoic acid | 0.9731 | 0.9852 | 6 | 1.92 | -0.2083 | -0.2984 | -0.2369 |
| Heptanoic acid | 0.9582 | 0.9853 | 7 | 2.41 | -0.1126 | -0.1868 | -0.122 |
| Octanoic acid | 0.9397 | 0.9852 | 8 | 3.05 | 0.0807 | -0.0753 | 0.0126 |
| Nonanoic acid | 0.9184 | 0.9853 | 9 | 3.47 | 0.3509 | 0.0363 | 0.1623 |
| Decanoic acid | 0.8986 | 0.9853 | 10 | 4.09 | 0.5063 | 0.1478 | 0.3039 |
| Undecanoic acid | 0.8813 | 0.9853 | 11 | 4.53 | 0.8983 | 0.2594 | 0.4319 |
| iso-Butyric acid | 0.9624 | 0.9834 | 4 | 0.60 | -0.3334 | -0.5216 | -0.2464 |
| Isovalerianic acid | 1.0071 | 0.9823 | 5 | 1.16 | -0.3415 | -0.41 | -0.4558 |
| Trimethylacetic acid | 0.9574 | 0.9819 | 5 | 1.47 | -0.2543 | -0.41 | -0.1853 |
| 3-Methylvaleric acid | 0.9657 | 0.9884 | 6 | 1.75 | -0.2331 | -0.2984 | -0.1966 |
| 4-Methylvaleric acid | 0.9964 | 0.9871 | 6 | 1.75 | -0.2724 | -0.2984 | -0.3637 |
| 2-Ethylbutyric acid | 0.9355 | 0.9854 | 6 | 1.68 | -0.1523 | -0.2984 | -0.0323 |
| 2-Propylpentanoic acid | 0.8905 | 0.9903 | 8 | 2.75 | 0.0258 | -0.0753 | 0.2803 |
| 2-Ethylhexanoic acid | 0.9122 | 0.9896 | 8 | 2.64 | 0.0756 | -0.0753 | 0.1622 |
| Succinic acid | 1.0511 | 0.9829 | 4 | -0.59 | -0.9395 | -0.5216 | -0.7291 |
| Glutaric acid | 1.0756 | 0.9839 | 5 | -0.29 | -0.6387 | -0.41 | -0.8286 |
| Adipic acid | 1.0345 | 0.9850 | 6 | 0.08 | -0.606 | -0.2984 | -0.5711 |
| Pimelic acid | 1.0336 | 0.9848 | 7 | 0.42 | -0.5845 | -0.1868 | -0.5323 |
| 3,3-Dimethylglutaric acid | 1.0614 | 0.9856 | 7 | 0.16 | -0.6643 | -0.1868 | -0.6837 |
| Suberic acid | 0.9991 | 0.9852 | 8 | 0.95 | -0.5116 | -0.0753 | -0.3107 |
| Sebacic acid | 0.9600 | 0.9853 | 10 | 2.01 | -0.2676 | 0.1478 | -0.0302 |
| 1,10-Decanedicarboxylic acid | 0.9181 | 0.9853 | 12 | 3.07 | -0.0863 | 0.3710 | 0.2655 |
| Crotonic acid | 1.0041 | 0.9462 | 4 | 0.72 | -0.5448 | -0.5216 | -0.4733 |
| trans-2-Pentenoic acid | 1.0254 | 0.9496 | 5 | 1.41 | -0.2774 | -0.41 | -0.5554 |
| trans-2-Hexenoic acid | 0.9961 | 0.9469 | 5 | 1.94 | -0.1279 | -0.41 | -0.3959 |
| ***Halogenated Acid*** | | | | | | | |
| 4-Bromobutyric acid | 0.6742 | 0.9994 | 4 | 0.68 | -0.7711 | -0.4453 | -0.6158 |
| 5-Bromovaleric acid | 0.6476 | 0.9992 | 5 | 1.21 | -0.6929 | -0.2197 | -0.5685 |
| 4-Chlorobutyric acid | 0.6786 | 0.9994 | 4 | 0.54 | -0.6773 | -0.4453 | -0.6075 |
| 3-Chloropropionic acid | 0.7333 | 0.9952 | 3 | 0.41 | -0.3321 | -0.671 | -0.6016 |
| 5-Chlorovaleric acid | 0.6419 | 0.9992 | 5 | 1.07 | -0.2857 | -0.2197 | -0.5793 |
| 2-Bromobutyric acid | 1.0508 | 0.9800 | 4 | 1.42 | 0.1221 | -0.4453 | 0.0971 |
| 2-Bromoisobutyric acid | 0.7178 | 0.9825 | 4 | 0.86 | -0.5845 | -0.4453 | -0.5333 |
| 2-Bromoisovaleric acid | 0.7562 | 0.9826 | 5 | 1.48 | -0.5492 | -0.2197 | -0.3629 |
| 2-Bromovaleric acid | 1.0422 | 0.9806 | 5 | 1.61 | -0.0423 | -0.2197 | 0.1785 |
| 2-Bromooctanoic acid | 1.0345 | 0.9806 | 8 | 3.19 | 0.4907 | 0.4574 | 0.4569 |
| 2-Bromohexanoic acid | 1.0382 | 0.9806 | 6 | 2.14 | 0.4547 | 0.0060 | 0.2686 |
| ***Mono Ester*** | | | | | | | |
| Ethyl acetate | 0.9420 | 0.9792 | 4 | 0.73 | -1.2968 | -1.3388 | -1.1201 |
| Propyl acetate | 0.9562 | 0.9799 | 5 | 1.24 | -1.2382 | -0.9743 | -1.0196 |
| Isopropyl acetate | 0.9664 | 0.9826 | 5 | 1.02 | -1.5900 | -0.9743 | -1.1309 |
| Butyl acetate | 0.9465 | 0.9801 | 6 | 1.78 | -0.4864 | -0.6098 | -0.6583 |
| Amyl acetate | 0.9408 | 0.9801 | 7 | 2.30 | 0.1625 | -0.2453 | -0.3407 |
| Hexyl acetate | 0.9328 | 0.9801 | 8 | 2.83 | -0.0087 | 0.1192 | 0.002 |
| Octyl acetate | 0.9115 | 0.9801 | 11 | 3.88 | 1.0570 | 1.2128 | 1.0007 |
| Decyl acetate | 0.8803 | 0.9801 | 12 | 4.94 | 1.8794 | 1.5773 | 1.5966 |
| Ethyl propionate | 0.9443 | 0.9821 | 5 | 1.21 | -0.9450 | -0.9743 | -0.8897 |
| Butyl propionate | 0.9379 | 0.9829 | 7 | 2.30 | 0.1704 | -0.2453 | -0.3091 |
| Isobutyl propionate | 0.9721 | 0.9837 | 7 | 2.17 | -0.6935 | -0.2453 | -0.6823 |
| Propyl propionate | 0.9502 | 0.9827 | 6 | 1.77 | -0.8148 | -0.6098 | -0.6987 |
| tert-Butyl propionate | 0.9288 | 0.9847 | 7 | 1.95 | -0.4095 | -0.2453 | -0.2098 |
| Ethyl butyrate | 0.9486 | 0.9879 | 6 | 1.77 | -0.4903 | -0.6098 | -0.6813 |
| Ethyl isobutyrate | 0.9406 | 0.9843 | 6 | 1.55 | -1.2709 | -0.6098 | -0.594 |
| Ethyl valerate | 0.9346 | 0.9889 | 7 | 2.30 | -0.3580 | -0.2453 | -0.2731 |
| Propyl butyrate | 0.9483 | 0.9886 | 7 | 2.30 | -0.4138 | -0.2453 | -0.4226 |
| Butyl butyrate | 0.9374 | 0.9887 | 8 | 2.83 | 0.5157 | 0.1192 | -0.0482 |
| Propyl valerate | 0.9381 | 0.9896 | 8 | 2.83 | 0.0094 | 0.1192 | -0.0558 |
| Amyl propionate | 0.9317 | 0.9829 | 5 | 2.83 | -0.0431 | -0.9743 | -0.7522 |
| Ethyl hexanoate | 0.9248 | 0.9891 | 6 | 2.83 | 0.0637 | -0.6098 | -0.4215 |
| Methyl butyrate | 0.9518 | 0.9832 | 5 | 1.29 | -1.2463 | -0.9743 | -0.9716 |
| Methyl valerate | 0.9380 | 0.9843 | 6 | 1.96 | -0.8448 | -0.6098 | -0.5656 |
| Methyl hexanoate | 0.9271 | 0.9845 | 7 | 2.30 | -0.5611 | -0.2453 | -0.1912 |
| Methyl heptanoate | 0.9157 | 0.9844 | 8 | 2.83 | 0.1039 | 0.1192 | 0.1886 |
| Methyl octanoate | 0.9027 | 0.9845 | 9 | 3.36 | 0.5358 | 0.4837 | 0.5859 |
| Methyl nonanoate | 0.8868 | 0.9845 | 10 | 3.88 | 1.0419 | 0.8482 | 1.0149 |
| Methyl decanoate | 0.8710 | 0.9845 | 11 | 4.41 | 1.3778 | 1.2128 | 1.4427 |
| Methyl undecanoate | 0.8562 | 0.9845 | 12 | 4.79 | 1.4248 | 1.5773 | 1.8596 |
| Methyl formate | 0.9611 | 0.8265 | 2 | 0.03 | -1.4982 | -2.0679 | -1.8393 |
| tert-Butyl formate | 0.9594 | 0.8379 | 5 | 0.97 | -1.3719 | -0.9743 | -1.0545 |
| ***Di Ester*** | | | | | | | |
| Diethyl malonate | 0.6983 | 1.0038 | 7 | 0.96 | -0.9975 | -0.8809 | -0.8413 |
| Diethyl sebacate | 0.5494 | 1.0009 | 14 | 3.90 | 1.3536 | 1.1221 | 1.2753 |
| Diethyl suberate | 0.5705 | 1.0010 | 12 | 2.84 | 0.7018 | 0.5498 | 0.7738 |
| Diethyl succinate | 0.6764 | 0.9996 | 8 | 1.19 | -0.8511 | -0.5948 | -0.5359 |
| Dimethyl malonate | 0.7367 | 0.9994 | 5 | -0.05 | -1.2869 | -1.4532 | -1.4261 |
| Dibutyl adipate | 0.5838 | 1.0017 | 14 | 3.90 | 0.7918 | 1.1221 | 1.1096 |
| Dimethyl succinate | 0.7085 | 0.9953 | 6 | 0.35 | -1.0573 | -1.1671 | -1.0904 |
| Diethyl adipate | 0.5991 | 1.0010 | 10 | 1.79 | -0.1265 | -0.0225 | 0.2362 |
| Dimethyl brassylate | 0.5361 | 0.9966 | 15 | 4.43 | 1.6536 | 1.4083 | 1.5392 |
| Dimethyl sebacate | 0.5703 | 0.9967 | 12 | 2.84 | 1.0106 | 0.5498 | 0.7748 |
| Dimethyl suberate | 0.5952 | 0.9967 | 10 | 1.79 | 0.2962 | -0.0225 | 0.255 |
| Diethyl pimelate | 0.5759 | 1.0006 | 11 | 2.31 | 0.4069 | 0.2636 | 0.5479 |
| Dibutyl suberate | 0.5574 | 1.0017 | 16 | 4.96 | 1.6556 | 1.6944 | 1.6366 |
| Diethyl butylmalonate | 0.6795 | 1.0140 | 11 | 3.02 | 0.5566 | 0.2636 | 0.0489 |
| Diethyl ethylmalonate | 0.6916 | 1.0133 | 9 | 1.96 | -0.2422 | -0.3086 | -0.4092 |
| Diethyl 3-oxopimelate | 0.7225 | 1.0050 | 11 | 1.49 | -0.3778 | 0.2636 | -0.1582 |
| Diethyl 4-oxopimelate | 0.7458 | 1.0002 | 11 | 1.54 | -0.6378 | 0.2636 | -0.2705 |
| Diethyl methylmalonate | 0.7004 | 1.0079 | 8 | 1.44 | -0.5114 | -0.5948 | -0.6515 |
| Diethyl propylmalonate | 0.6837 | 1.0140 | 10 | 2.49 | 0.1341 | -0.0225 | -0.1713 |
| Dibutyl succinate | 0.6602 | 1.0003 | 12 | 3.60 | 0.5123 | 0.5498 | 0.3418 |
| ***Aldehyde*** | | | | | | | |
| Propionaldehyde | 0.8905 | 0.5592 | 3 | 0.59 | -0.4855 | -0.7336 | -0.5798 |
| Butyraldehyde | 0.8722 | 0.5638 | 4 | 0.88 | -0.3805 | -0.5106 | -0.4108 |
| Isobutyraldehyde | 0.9295 | 0.5589 | 4 | 0.61 | -0.4328 | -0.5106 | -0.5555 |
| Valeraldehyde | 0.8620 | 0.5651 | 5 | 1.36 | -0.0223 | -0.2876 | -0.2623 |
| 2-Methyl-butyraldehyde | 0.8494 | 0.5644 | 5 | 1.14 | -0.3107 | -0.2876 | -0.2305 |
| Hexylaldehyde | 0.8384 | 0.5624 | 6 | 1.78 | -0.1731 | -0.0646 | -0.0799 |
| 2-Methylvaleraldehyde | 0.8354 | 0.5655 | 6 | 1.67 | -0.4745 | -0.0646 | -0.0723 |
| 2-Ethylbutyraldehyde | 0.8429 | 0.5542 | 6 | 1.67 | -0.0544 | -0.0646 | -0.0913 |
| 3,3-Dimethylbutyraldehyde | 0.9114 | 0.5579 | 6 | 1.63 | -0.3744 | -0.0646 | -0.2642 |
| Heptaldehyde | 0.8517 | 0.5653 | 7 | 2.42 | -0.0019 | 0.1584 | 0.0093 |
| 2-Ethylhexanal | 0.8268 | 0.5555 | 8 | 2.73 | 0.1608 | 0.3814 | 0.1949 |
| trans-4-Decen-1-al | 0.6717 | 0.5642 | 10 | 4.05 | 1.2076 | 0.8275 | 0.832 |
| cis-7-Decen-1-al | 0.5588 | 0.5652 | 10 | 3.52 | 0.9485 | 0.8275 | 1.1171 |
| ***Ketones*** | | | | | | | |
| Acetone | 0.8709 | 0.6969 | 3 | -0.24 | -2.2036 | -2.203 | -2.2784 |
| 2-Butanone | 0.8544 | 0.7020 | 4 | 0.29 | -1.7457 | -1.7884 | -1.8354 |
| 2-Pentanone | 0.8175 | 0.7072 | 5 | 0.91 | -1.2224 | -1.3737 | -1.3268 |
| 3-Pentanone | 0.8315 | 0.7033 | 5 | 0.85 | -1.4561 | -1.3737 | -1.3719 |
| 4-Methyl-2-pentanone | 0.8315 | 0.7060 | 6 | 1.31 | -1.2085 | -0.959 | -0.982 |
| 2-Heptanone | 0.7975 | 0.7084 | 7 | 1.98 | -0.4872 | -0.5444 | -0.4827 |
| 5-Methyl-2-hexanone | 0.8053 | 0.7105 | 7 | 1.88 | -0.6459 | -0.5444 | -0.5078 |
| 4-Heptanone | 0.8108 | 0.7151 | 7 | 1.91 | -0.6690 | -0.5444 | -0.5255 |
| 2-Octanone | 0.7948 | 0.7085 | 8 | 2.37 | -0.1455 | -0.1297 | -0.0841 |
| 2-Nonanone | 0.7926 | 0.7085 | 9 | 3.14 | 0.6598 | 0.2849 | 0.3129 |
| 2-Decanone | 0.7912 | 0.7085 | 10 | 3.73 | 0.5822 | 0.6996 | 0.7072 |
| 3-Decanone | 0.7992 | 0.7171 | 10 | 3.49 | 0.6265 | 0.6996 | 0.6815 |
| 2-Undecanone | 0.7901 | 0.7085 | 11 | 4.09 | 1.5346 | 1.1142 | 1.1007 |
| 2-Dodecanone | 0.7893 | 0.7085 | 12 | 4.55 | 1.6696 | 1.5289 | 1.4931 |
| 7-Tridecanone | 0.7811 | 0.7177 | 13 | 5.08 | 1.5214 | 1.9435 | 1.9094 |

* Taken from reference 10.

**Table 2**: Electrophilicity (ω), maximum atomic charge ( $Q_k^{\max}$ ), number of carbon atoms (Nc), log P along with the experimental and calculated values of log ($IGC_{50}^{-1}$) for the complete set of aliphatic donor compounds with *Tetrahymena pyriformis*.

| Molecules | ω | $Q_k^{\max}$ | Nc | log P* | pIGC$_{50}$ Expt.* | Calc. (Nc) | Calc. (Nc, ω) |
|---|---|---|---|---|---|---|---|
| ***Amino Alcohol*** | | | | | | | |
| 2-(methylamino)ethanol | 0.5611 | -0.7684 | 3 | -0.94 | -1.8202 | -1.6530 | -1.961 |
| 4-amino-1-butanol | 0.6562 | -0.8278 | 4 | -1.06 | -0.9752 | -1.4275 | -0.9598 |
| 2-(ethylamino)ethanol | 0.5658 | -0.7656 | 4 | -0.46 | -1.6491 | -1.4275 | -1.7723 |
| 2-Propylaminoethanol | 0.5548 | -0.7657 | 5 | 0.07 | -1.6842 | -1.2020 | -1.7248 |
| DL-2-amino-1-pentanol | 0.6623 | -0.8457 | 5 | 0.07 | -0.6718 | -1.2020 | -0.7586 |
| 3-amino-2,2-dimethyl-1-propanol | 0.6792 | -0.8558 | 5 | -0.79 | -0.9246 | -1.2020 | -0.6067 |
| 6-amino-1-hexanol | 0.6297 | -0.8512 | 6 | -0.01 | -0.958 | -0.9764 | -0.9052 |
| DL-2-amino-1-hexanol | 0.6621 | -0.8458 | 6 | 0.60 | -0.5848 | -0.9764 | -0.614 |
| DL-2-amino-3-methyl-1-butanol | 0.6306 | -0.8569 | 5 | -0.06 | -0.5852 | -1.2020 | -1.0435 |
| 2-amino-3,3-dimethyl-butanol | 0.6430 | -0.8599 | 6 | 0.34 | -0.7178 | -0.9764 | -0.7857 |
| 2-amino-3-methyl-1-pentanol | 0.6325 | -0.8607 | 6 | 0.47 | -0.6594 | -0.9764 | -0.88 |
| 2-amino-4-methyl-pentanol | 0.6484 | -0.8574 | 6 | 0.47 | -0.6191 | -0.9764 | -0.7371 |
| 2-(tert.butylamino)ethanol | 0.5856 | -0.7671 | 6 | 0.41 | -1.673 | -0.9764 | -1.3016 |
| Diethanolamine | 0.5880 | -0.7685 | 4 | -1.43 | -1.7941 | -1.4275 | -1.5728 |
| 1,3-diamino-2-hydroxy-propane | 0.6407 | -0.8517 | 3 | -2.05 | -1.4275 | -1.6530 | -1.2456 |
| N-methyldiethanol amine | 0.5309 | -0.7675 | 5 | -1.04 | -1.8338 | -1.2020 | -1.9396 |
| 3-(methylamino)-1,2-propanediol | 0.5936 | -0.7897 | 4 | -1.82 | -1.5341 | -1.4275 | -1.5225 |
| Triethanolamine | 0.5602 | -0.7678 | 6 | -1.00 | -1.7488 | -0.9764 | -1.5298 |
| ***Acetylenic Alcohols*** | | | | | | | |
| 3-Butyn-2-ol | 0.7438 | -0.7525 | 4 | 0.14 | -0.4024 | -0.8795 | -0.781 |
| 1-Pentyn-3-ol | 0.7443 | -0.7565 | 5 | 0.67 | -1.1776 | -0.5463 | -0.4085 |
| 2-Pentyn-1-ol | 0.6737 | -0.7387 | 5 | 0.89 | -0.5724 | -0.5463 | -0.6729 |
| 2-Penten-4-yn-1-ol | 0.6042 | -0.7593 | 6 | -0.01 | -0.5549 | -0.2130 | -0.5625 |
| 1-Hexyn-3-ol | 0.7265 | -0.7565 | 6 | 1.2 | 0.6574 | -0.2130 | -0.1044 |
| 1-Heptyn-3-ol | 0.7227 | -0.7566 | 7 | 1.73 | -0.265 | 0.1202 | 0.252 |
| 4-Heptyn-3-ol | 0.6704 | -0.7601 | 7 | 1.73 | -0.0336 | 0.1202 | 0.0561 |
| 2-Octyn-1-ol | 0.6495 | -0.7388 | 8 | 2.48 | 0.1944 | 0.4534 | 0.3485 |
| 2-Nonyn-1-ol | 0.6487 | -0.7388 | 9 | 3.01 | 0.6486 | 0.7867 | 0.7162 |
| 2-Decyn-1-ol | 0.6481 | -0.7388 | 10 | 3.54 | 0.9855 | 1.1199 | 1.0847 |
| 2-Tridecyn-1-ol | 0.6474 | -0.7388 | 13 | 5.13 | 2.3665 | 2.1196 | 2.1941 |
| 4-Methyl-1-pentyn-3-ol | 0.7265 | -0.7565 | 6 | 1.07 | -0.0267 | -0.2130 | -0.1044 |
| 4-Methyl-1-heptyn-3-ol | 0.7018 | -0.7566 | 8 | 2.13 | 0.7426 | 0.4534 | 0.5444 |
| ***Unsaturated Alcohol*** | | | | | | | |
| 2-methyl-3-buten-2-ol | 0.6225 | -0.7821 | 5 | 0.52 | -1.3889 | -1.3007 | -1.2972 |
| 4-pentyn-1-ol | 0.7562 | -0.7561 | 5 | -0.01 | -1.4204 | -1.3007 | -1.5947 |
| 2-methyl-3-butyn-2-ol | 0.7465 | -0.7635 | 5 | 0.28 | -1.3114 | -1.3007 | -1.5731 |
| trans-3-hexen-1-ol | 0.4768 | -0.7625 | 6 | 1.40 | -0.7772 | -0.8914 | -0.6088 |
| cis-3-hexen-1-ol | 0.5049 | -0.7703 | 6 | 1.40 | -0.8091 | -0.8914 | -0.6714 |
| 5-hexyn-1-ol | 0.7024 | -0.7667 | 6 | 0.52 | -1.2948 | -0.8914 | -1.1108 |
| 3-methyl-1-pentyn-3-ol | 0.7596 | -0.768 | 6 | 1.07 | -1.3226 | -0.8914 | -1.2381 |
| 4-hexen-1-ol | 0.4780 | -0.7634 | 6 | 1.40 | -0.754 | -0.8914 | -0.6115 |
| 5-hexen-1-ol | 0.5493 | -0.7636 | 6 | 1.40 | -0.8411 | -0.8914 | -0.7702 |
| 4-pentyn-2-ol | 0.7275 | -0.772 | 5 | 0.12 | -1.6324 | -1.3007 | -1.5308 |
| 5-hexyn-3-ol | 0.7249 | -0.7808 | 6 | 0.65 | -1.4043 | -0.8914 | -1.1609 |
| 3-heptyn-1-ol | 0.6046 | -0.7656 | 7 | 1.40 | -0.3231 | -0.4820 | -0.5291 |
| 4-heptyn-2-ol | 0.6054 | -0.7723 | 7 | 1.18 | -0.616 | -0.4820 | -0.5309 |
| 3-octyn-1-ol | 0.5983 | -0.7656 | 8 | 1.93 | 0.017 | -0.0727 | -0.1509 |
| 3-nonyn-1-ol | 0.5942 | -0.7656 | 9 | 2.46 | 0.3401 | 0.3366 | 0.2223 |
| 2-propen-1-ol | 0.6632 | -0.7531 | 3 | 0.17 | -1.9178 | -2.1193 | -2.116 |
| 2-buten-1-ol | 0.5471 | -0.7588 | 4 | 0.34 | -1.4719 | -1.7100 | -1.4935 |
| (+/-)-3-buten-2-ol | 0.6295 | -0.771 | 4 | 0.12 | -1.0529 | -1.7100 | -1.6769 |
| cis-2-buten-1,4-diol | 0.6479 | -0.7578 | 5 | -0.81 | -2.1495 | -1.3007 | -1.3537 |
| cis-2-penten-1-ol | 0.5885 | -0.755 | 5 | 0.87 | -1.1052 | -1.3007 | -1.2215 |
| 3-penten-2-ol | 0.5738 | -0.7709 | 5 | 0.65 | -1.401 | -1.3007 | -1.1888 |
| trans-2-hexen-1-ol | 0.4642 | -0.7591 | 6 | 1.40 | -0.4718 | -0.8914 | -0.5808 |
| 1-hexen-3-ol | 0.6304 | -0.7748 | 6 | 1.18 | -0.8113 | -0.8914 | -0.9506 |
| cis-2-hexen-1-ol | 0.5381 | -0.7588 | 6 | 1.40 | -0.7767 | -0.8914 | -0.7452 |
| trans-2-octen-1-ol | 0.4621 | -0.759 | 8 | 2.45 | 0.3654 | -0.0727 | 0.1521 |
| ***Amines*** | | | | | | | |
| Propylamine | 0.6353 | -0.8330 | 3 | 0.47 | -0.7075 | -1.005 | -1.003 |
| Butylamine | 0.6334 | -0.8325 | 4 | 0.97 | -0.5735 | -0.891 | -0.891 |
| N-Methylpropylamine | 0.5455 | -0.6865 | 4 | 0.84 | -0.8087 | -0.891 | -0.876 |
| Amylamine | 0.6218 | -0.8510 | 5 | 1.49 | -0.4848 | -0.778 | -0.777 |
| N-Methylbutylamine | 0.5416 | -0.6826 | 5 | 1.33 | -0.6784 | -0.778 | -0.764 |
| Hexylamine | 0.6213 | -0.8510 | 6 | 2.06 | -0.2197 | -0.664 | -0.666 |
| Isopropylamine | 0.6842 | -0.8479 | 3 | 0.26 | -0.8635 | -1.005 | -1.011 |
| Isobutylamine | 0.6703 | -0.8631 | 4 | 0.73 | -0.2616 | -0.891 | -0.897 |
| N,N-Dimethylethylamine | 0.4764 | -0.5751 | 4 | 0.70 | -0.9083 | -0.891 | -0.864 |
| (+/-)-sec-Butylamine | 0.6626 | -0.8473 | 4 | 0.74 | -0.6708 | -0.891 | -0.896 |
| Isoamylamine | 0.6505 | -0.8314 | 5 | 1.32 | -0.5774 | -0.778 | -0.782 |
| 1-Methylbutylamine | 0.6543 | -0.8469 | 5 | 1.23 | -0.6846 | -0.778 | -0.783 |
| 1-Ethylpropylamine | 0.6303 | -0.8455 | 7 | 1.23 | -0.8129 | -0.551 | -0.555 |
| 2-Methylbutylamine | 0.6449 | -0.8380 | 5 | 1.32 | -0.4774 | -0.778 | -0.781 |
| N,N-Diethylmethylamine | 0.4888 | -0.5714 | 5 | 0.95 | -0.7559 | -0.778 | -0.755 |
| tert-Butylamine | 0.7163 | -0.8541 | 4 | 0.40 | -0.8973 | -0.891 | -0.905 |
| tert-Amylamine | 0.6995 | -0.8592 | 5 | 1.10 | -0.6978 | -0.778 | -0.790 |
| (+/-)-1,2-Dimethylpropylamine | 0.6367 | -0.8457 | 5 | 1.10 | -0.7095 | -0.778 | -0.780 |
| Propargylamine | 0.6898 | -0.8084 | 3 | -0.43 | -0.826 | -1.005 | -1.012 |
| N-Methylpropargylamine | 0.6355 | -0.6632 | 4 | 0.08 | -0.9818 | -0.891 | -0.891 |
| 1-Dimethylamino-2-propyne | 0.5750 | -0.5392 | 5 | -0.01 | -1.1451 | -0.778 | -0.769 |
| 1,1-Dimethylpropargylamine | 0.6681 | -0.8289 | 5 | 0.64 | -0.9104 | -0.778 | -0.785 |
| 2-Methoxyethylamine | 0.6585 | -0.8568 | 3 | -0.67 | -1.7903 | -1.005 | -1.007 |
| 3-Methoxypropylamine | 0.6608 | -0.8478 | 4 | -1.02 | -1.7725 | -0.891 | -0.895 |
| 3-Ethoxypropylamine | 0.6592 | -0.8479 | 5 | -0.49 | -1.7027 | -0.778 | -0.784 |

* Taken from reference 10.

**Table 3.** Regression models for different groups of aliphatic compounds for estimating their toxicity towards *Tetrahymena pyriformis*

| Molecules | Regression Equations | R | SD |
|---|---|---|---|
| *Aliphatic Electron Acceptors* | | | |
| **Diols** | $pIGC_{50} = 0.4497 \times N_C - 4.0855$ | 0.9683 | 0.2781 |
| **(N=10)** | $pIGC_{50} = -12.4224 \times \omega + 0.3554 \times N_C + 7.6094$ | 0.9826 | 0.2070 |
| **Halogenated Alcohols** | $pIGC_{50} = 0.3271 \times N_C - 2.0248$ | 0.8923 | 0.3852 |
| **(N=11)** | $pIGC_{50} = - 6.2863 \times \omega + 0.1982 \times N_C + 4.5793$ | 0.9424 | 0.2855 |
| **Saturated Alcohols** | $pIGC_{50} = 0.4144 \times N_C - 3.0801$ | 0.9634 | 0.3456 |
| **(N=32)** | $pIGC_{50} = 0.8927 \times \omega + 0.4261 \times N_C - 3.9484$ | 0.9636 | 0.3451 |
| **Carboxylic Acids** | $pIGC_{50} = 0.1116 \times N_C - 0.9678$ | 0.6676 | 0.2917 |
| **(N=28)** | $pIGC_{50} = - 5.4426 \times \omega + 0.0338 \times N_C + 4.8562$ | 0.8801 | 0.1860 |
| **Halogenated Acids** | $pIGC_{50} = 0.2257 \times N_C - 1.3481$ | 0.6564 | 0.3632 |
| **(N=11)** | $pIGC_{50} = 1.8930 \times \omega + 0.0976 \times N_C - 2.2827$ | 0.9186 | 0.1903 |
| **Mono Esters** | $pIGC_{50} = 0.3645 \times N_C - 2.7969$ | 0.9189 | 0.3710 |
| **(N= 31)** | $pIGC_{50} = -10.9131 \times \omega + 0.2554 \times N_C + 8.1384$ | 0.9352 | 0.3330 |
| **Di Esters** | $pIGC_{50} = 0.2861 \times N_C - 2.884$ | 0.9299 | 0.3382 |
| **(N=20)** | $pIGC_{50} = -4.8166 \times \omega + 0.1999 \times N_C + 1.1227$ | 0.9636 | 0.2460 |
| **Aldehydes** | $pIGC_{50} = 0.2230 \times N_C - 1.4027$ | 0.8980 | 0.2459 |
| **(N=13)** | $pIGC_{50} = -2.5248 \times \omega + 0.1228 \times N_C + 1.3002$ | 0.9332 | 0.2008 |
| **Ketones** | $pIGC_{50} = 0.4147 \times N_C - 3.4470$ | 0.9850 | 0.2249 |
| **(N=15)** | $pIGC_{50} = -3.2176 \times \omega + 0.38989 \times N_C - 0.6459$ | 0.9855 | 0.2211 |
| *Aliphatic Electron Donors* | | | |
| **Amino Alcohols** | $pIGC_{50} = 0.2255 \times N_C - 2.3296$ | 0.4711 | 0.4596 |
| **(N=18)** | $pIGC_{50} = 8.9875 \times \omega + 0.1464 \times N_C - 7.4431$ | 0.9152 | 0.2100 |
| **Acetylenic Alcohols** | $pIGC_{50} = 0.3332 \times N_C - 2.2125$ | 0.8942 | 0.4218 |
| **(N=13)** | $pIGC_{50} = 3.7452 \times \omega + 0.3707 \times N_C - 5.0494$ | 0.9080 | 0.3947 |
| **Unsaturated Alcohols** | $pIGC_{50} = 0.4093 \times N_C - 3.3473$ | 0.8580 | 0.3311 |
| **(N=25)** | $pIGC_{50} = -2.2250 \times \omega + 0.3641 \times N_C - 1.7327$ | 0.9136 | 0.2622 |
| **Amines** | $pIGC_{50} = 0.1136 \times N_C - 1.3456$ | 0.2711 | 0.3965 |
| **(N=25)** | $pIGC_{50} = -0.1700 \times \omega + 0.1116 \times N_C - 1.2295$ | 0.2723 | 0.3964 |
| | $pIGC_{50} = 0.1162 \times Q_k^{max} + 2.1524 \times \omega^{-}_{max} + 0.0669 \times N_C - 1.8782$ | 0.8692 | 0.2037 |

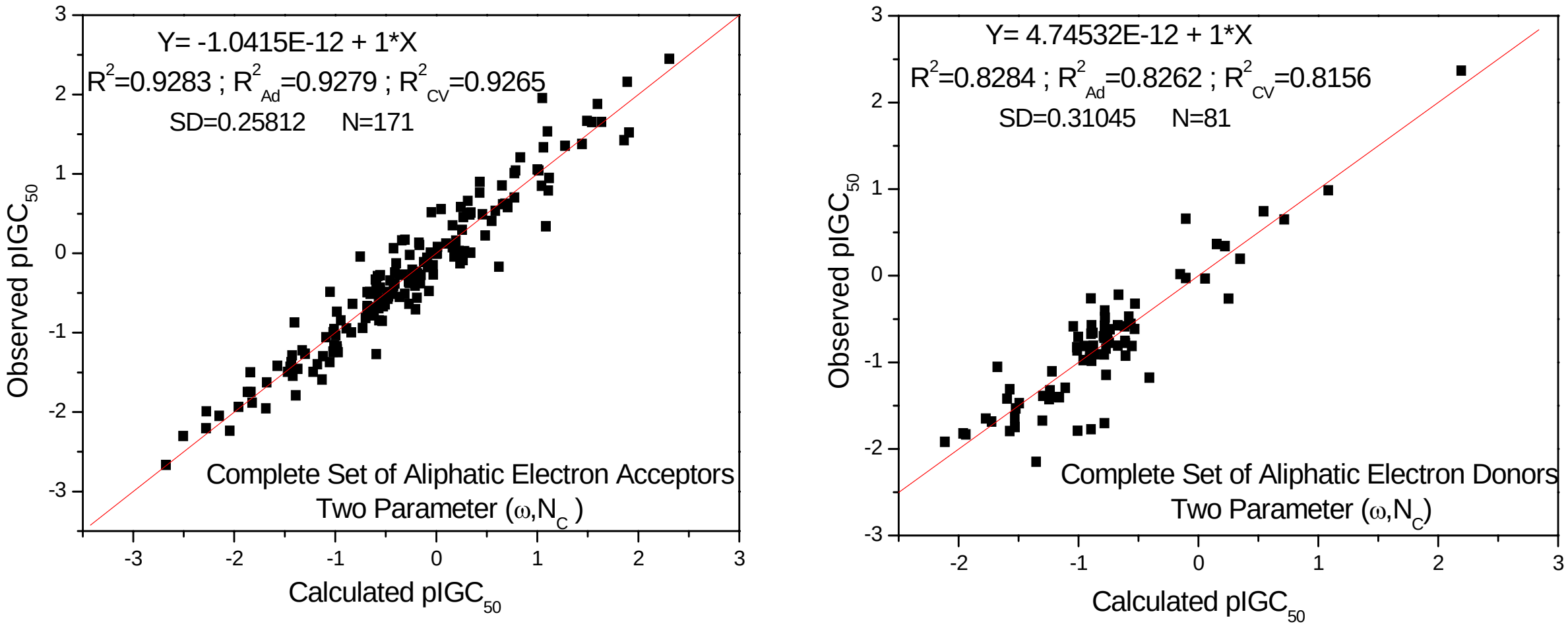

**(a)** **(b)**

**Figure 1.** Observed versus calculated $pIGC_{50}$ values using two-parameter ($\omega$, $N_C$) regression model for the (a) Complete set of aliphatic electron acceptors and (b) Complete set of aliphatic electron donors. Please see the text for details.

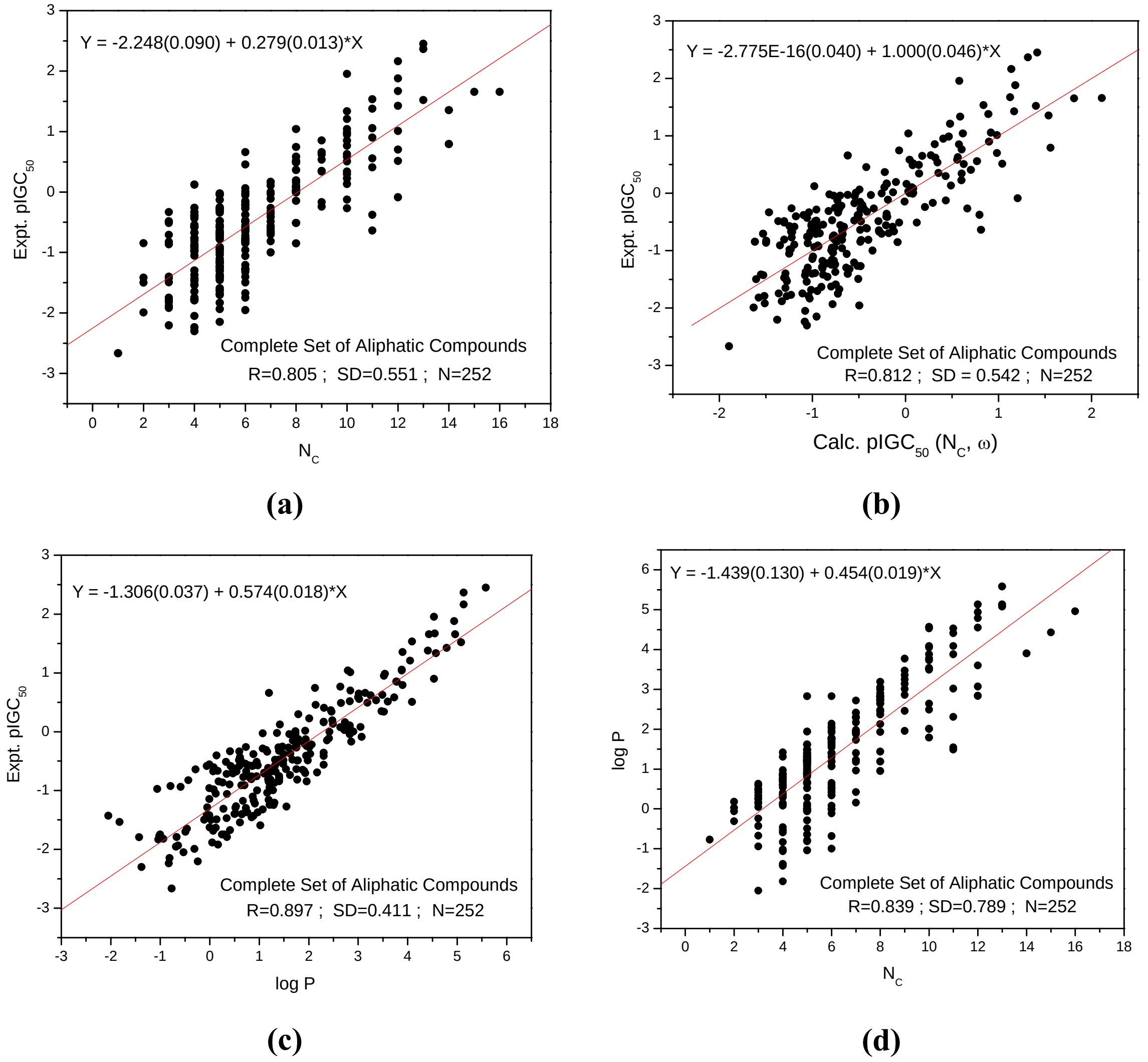

**Figure 2.** Observed $pIGC_{50}$ versus the a) Number of carbon atoms ($N_C$), b) Calculated $pIGC_{50}$ values using two-parameter ($\omega$, $N_C$) regression model and (c) log P along with the (d) inter-correlation between log P and $N_C$ for the complete set of 252 aliphatic compounds.

Graphical Abstract

# New QSTR Models for the Toxicity Analysis†

P. K. Chattaraj[a,*], D. R. Roy[a], S. Giri[a], S. Mukherjee[a], V. Subramanian[b,*], P. Bultinck[c,*] and S. Van Damme[c]

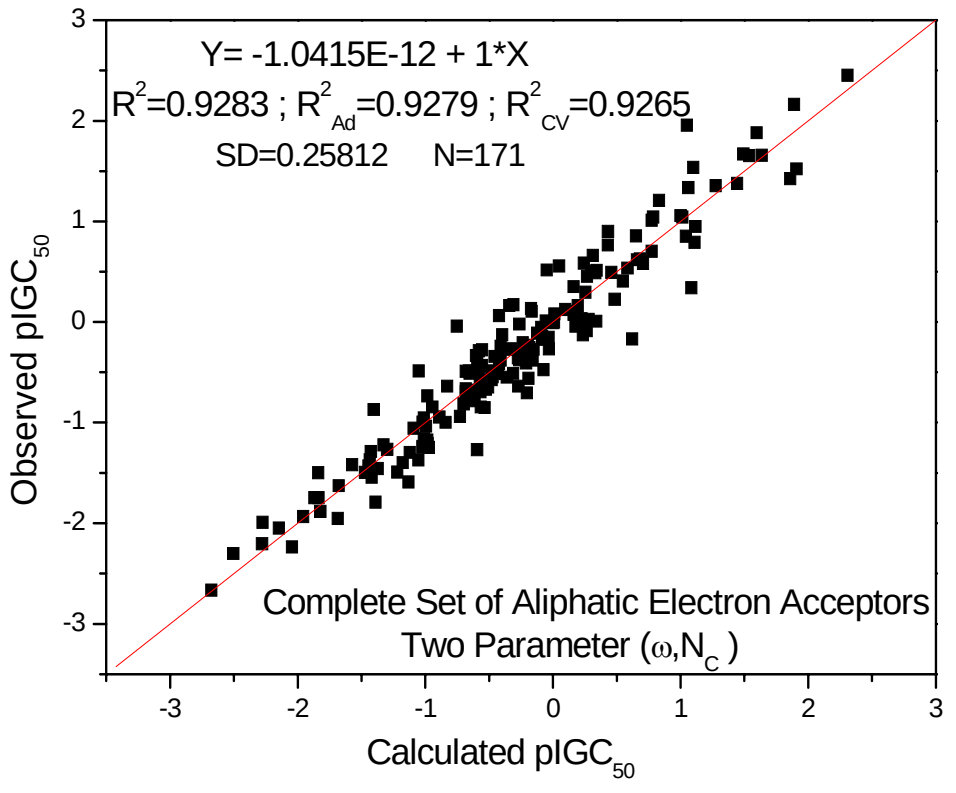

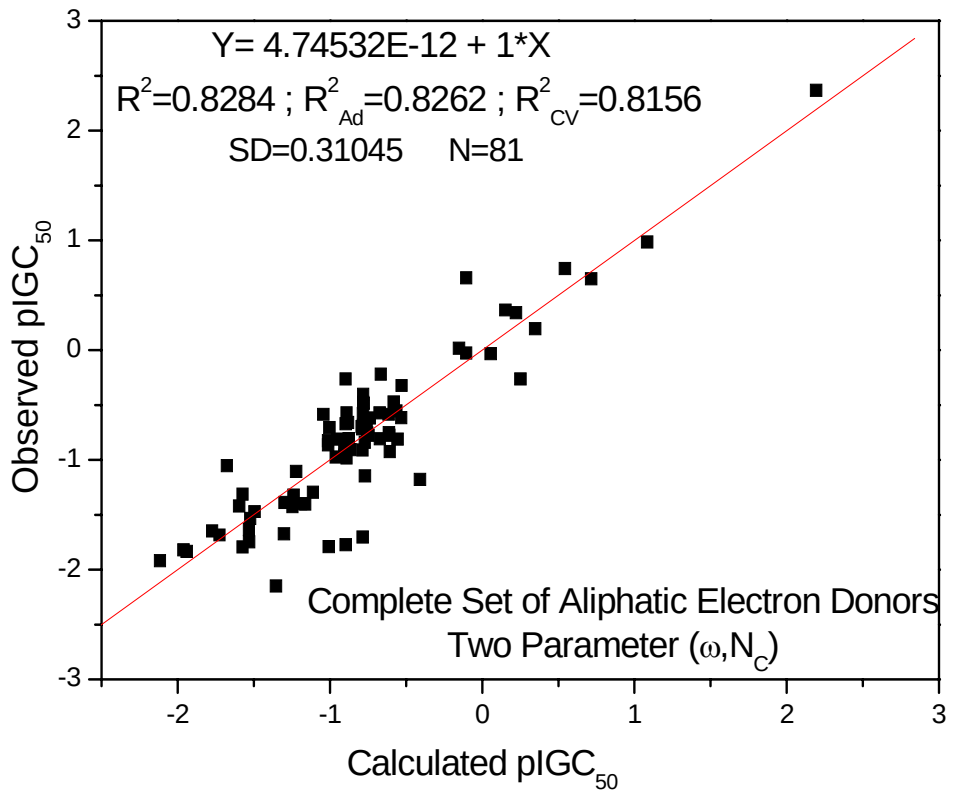

A deceptively simple descriptor, viz. the number of carbon / non-hydrogenic atoms present in a molecule is prescribed for the development of useful quantitative-structure-toxicity-relationship (QSTR) models. Please see the text for details.